\UseRawInputEncoding 
\documentclass{vgtc}                          
\ifpdf
  \pdfoutput=1\relax                   
  \usepackage{graphicx}                
  \DeclareGraphicsExtensions{.pdf,.png,.jpg,.jpeg} 
\else
  \usepackage{graphicx}                
  \DeclareGraphicsExtensions{.eps}     
\fi%

\graphicspath{{figures/}{pictures/}{images/}{./}} 
\usepackage{soul}
\usepackage{microtype}                 
\PassOptionsToPackage{warn}{textcomp}  
\usepackage{textcomp}                  
\usepackage{mathptmx}                  
\usepackage{times}                     
\usepackage{cite}                      
\usepackage{tabu}                      
\usepackage{booktabs}                  
\usepackage{dingbat}
\usepackage[table]{xcolor}
\usepackage{gensymb}
\definecolor{mycolor1}{HTML}{f0f9e8}
\definecolor{mycolor2}{HTML}{bae4bc}
\definecolor{mycolor3}{HTML}{7bccc4}

\onlineid{0}

\vgtccategory{Research}

\vgtcinsertpkg

\title{An Exploration of Hands-free Text Selection for Virtual Reality Head-Mounted Displays}

\author{Xuanru Meng\thanks{e-mail: xuanru.meng18@student.xjtlu.edu.cn}\\ 
             \parbox{1.7in}{\scriptsize \centering Xi'an Jiaotong-Liverpool University\\ Suzhou, China} %
\and Wenge Xu\thanks{e-mail: wenge.xu@bcu.ac.uk}\\ %
          \parbox{1.8in}{\scriptsize \centering DMT Lab, Birmingham City University\\ Birmingham, UK} %
\and Hai-Ning Liang\thanks{e-mail: haining.liang@xjtlu.edu.cn (\textit{corresponding author})}\\ %
     \parbox{1.7in}{\scriptsize \centering Xi'an Jiaotong-Liverpool University\\ Suzhou, China}}

\abstract{Hand-based interaction, such as using a handheld controller or making hand gestures, has been widely adopted as the primary method for interacting with both virtual reality (VR) and augmented reality (AR) head-mounted displays (HMDs). In contrast, hands-free interaction avoids the need for users' hands and although it can afford additional benefits, there has been limited research in exploring and evaluating hands-free techniques for these HMDs. As VR HMDs become ubiquitous, people will need to do text editing, which requires selecting text segments. Similar to hands-free interaction, text selection is underexplored. This research focuses on both, text selection via hands-free interaction. Our exploration involves a user study with 24 participants to investigate the performance, user experience, and workload of three hands-free selection mechanisms (Dwell, Blink, Voice) to complement head-based pointing. Results indicate that Blink outperforms Dwell and Voice in completion time. Users' subjective feedback also shows that Blink is the preferred technique for text selection. This work is the first to explore hands-free interaction for text selection in VR HMDs. Our results provide a solid platform for further research in this important area.%
} 


\CCScatlist{
  \CCScatTwelve{Human-centered computing}{Human computer interaction (HCI)}{Interaction paradigms}{Virtual reality};
  \CCScatTwelve{Human-centered computing}{Human computer interaction (HCI)}{interaction techniques}{};
  \CCScatTwelve{Human-centered computing}{Interaction design}{Empirical studies in interaction design}{}
}

\begin{document}


\firstsection{Introduction}

\maketitle


Today's virtual reality (VR) and augmented reality (AR) head-mounted displays (HMDs) predominately prioritize hand-based interaction via a handheld controller or hand/finger gestures. Despite its  functional advantages and wide adoption, relying on hands for interaction can be impractical and at times impossible in many task scenarios, for instance, in manual assembly and manufacturing tasks \cite{baird_evaluating_1999, 183317}, emergency responses \cite{4343891}, text entry activities \cite{hands_free_text_input, Lu.2021.Itext} and many others \cite{729527,10.1145/2702123.2702305}. In addition, users who have hand/arm impairment are unlikely to be able to use their hands to hold a controller or perform hand gestures accurately. An efficient and usable hands-free interaction method would be the most convenient and practical solution in scenarios where hands-based interaction is impractical. There have been some attempts to explore hands-free interaction for HMDs for text entry \cite{hands_free_text_input,Lu.2021.Itext}, system control \cite{xu.2019.dmove, DepthMove}, and rapid activation of glanceable objects \cite{glanceable_feiyu}. However, there is very limited attention to text selection tasks for VR HMDs.

Text selection is an essential task when reading text content such as newspapers, magazines, and academic papers to highlight important elements for later reference or to copy/cut and transfer the content to another document, application, or platform. It can also be useful when coding in immersive environments with an unlimited screen size. While there has been some work on text selection in HMDs, it is not based on hands-free interaction. For instance, EYEditor \cite{ghosh_eyeditor_2020} uses a ring mouse for cursor navigation and text selection, where a button is used for placing the cursor before and after a text fragment to be selected while the selection is made via a touchpad. Lee et al. \cite{lee_one-thumb_2020} have employed a force-sensitive smartphone as their input device, where users exert a force on a thumb-sized circular button to select the desired text fragment. Similarly, Darbar et al. \cite{darbar_exploring_2021} have explored the use of a smartphone as the input mechanism for text selection in AR HMDs. They found that continuous touch is more efficient than discrete touch, spatial movement, and ray casting. These methods all employed an external handheld device for accomplishing text selection. In this research, we are interested in evaluating and comparing hands-free interaction methods for text selection in HMDs. As described earlier, hands-free interaction is helpful in many scenarios where hands or handheld controllers are not available or impractical to use.

Interacting with virtual content in HMDs usually requires (1) a pointing mechanism for the identification of the objects to be selected priors to interact with them \cite{xu_pointing_2019,yu.2018.targetselection}, and (2) a selection mechanism (e.g., signal/command/action) to indicate the selection \cite{Mine95virtualenvironment,yu.2021.gaze}. In this study, we focus on head-based pointing as our primary pointing method as it is a mature and cost-effective way to control a cursor and has been widely adopted as a standard way for pointing at virtual objects in HMDs for hands-free interaction \cite{pinpointing}. Studies have demonstrated that head-based pointing is accurate, comfortable, and convenient \cite{pinpointing,6970627,lei.2021.pointing}. However, head-based pointing lacks an intrinsic mechanism to confirm a highlighted/identified selection \cite{ESTEVES2020102414}. To enable selection with head-based pointing, this work first explores potential selection mechanisms (including dwell, eye blinks, voice (HumHum: first-letter-hum plus last-latter-hum, and Hummer: continuous humming), neck forward/backward motions) that are available in the literature and can be used for text selection in HMDs. It then, through pilot studies and a set of three usability criteria, narrows them down into three suitable candidates (dwell, eye blinks, voice-hummer) for the final experiment. Our results with 24 participants suggest that eye blinking is the best hands-free selection candidate as it has the fastest performance, best accuracy, highest experience, and lowest workload. 

The main contribution of this work is a first formal evaluation of three hands-free text selection mechanisms for VR HMDs in terms of their performance, user experience, and workload. While a recent paper has explored text selection techniques in VR \cite{xu_withhands_2022}, their techniques are non-hands-free and are based on hand-based and controller-based interaction. As such, our work can serve as the foundation for further research linking text selection and hands-free interaction.

\section{Evaluated Selection Methods}
In this section, we described the selected hands-free methods evaluated in this work. They were developed in Unity (v2019.4.10f1).

\subsection{Selection Methods} \label{section:SelectionMethods}
We implemented and tested the following methods in pilot trials to determine their suitability for the main experiment. 

\subsubsection{Dwell}
The most commonly used hands-free selection method is the dwell technique \cite{Dwell_Technique}. It was originally developed to avoid the effects of eye-tracking jitters for eye gaze interaction and has also been widely tested in various 3D interaction tasks, e.g., text entry \cite{hands_free_text_input,tap_dwel_gesture,Ringtext}, rapid activation \cite{glanceable_feiyu}, and system control \cite{DepthMove}. Therefore, we have included this technique as the baseline technique. With the Dwell technique (Dwell for short), the user starts the selection of the desired text fragment by dwelling (staying or hovering the pointer on an area for 1s) at the beginning of the first letter and ends the selection by dwelling (1s) again at the end of the text fragment. Determining the dwell time is important because if it is too long, it will make interaction inefficient unnecessarily but if it is too short it can lead to high errors and a stressful interaction. We first checked prior studies (from text entry in AR/VR, for example, \cite{grubert2018text,Ringtext,Marco_dwell, Lu.2021.Itext}). It ranges from 400ms to 1s. We run some pilot tests with several possibilities within this range and found that 1s is the most optimal dwell time.

\subsubsection{Eye Blinks}
Eye blinking has been explored and used frequently in assistive technologies \cite{blink_selection,grauman_communication_2003} and has been lately incorporated into HMDs for text entry \cite{hands_free_text_input} and rapid activation of glanceable information \cite{glanceable_feiyu}. Lu et al. \cite{hands_free_text_input} found that using eye blinks (of both eyes) outperforms Dwell in text entry tasks in VR regarding performance and experience, while Lu et al. \cite{glanceable_feiyu} suggested that blinking is preferred by users when accessing information in the real world in AR. 

Although eye blinks (Blinking for short) is a promising selection mechanism, it requires add-on or built-in eye trackers for HMDs, which could be an issue with earlier generations of HMDs. However, manufacturers are now integrating the eye-trackers with their HMDs (e.g., HTC VIVE Pro Eye, Pico Neo 2 Eye, FOVE, HoloLens 2, Magic Leap 1, and LooxidVR), and there is now increasing interest in using eye data for interactive operations \cite{yu.2021.gaze}, for example, Blinking for confirming a selection \cite{hands_free_text_input,glanceable_feiyu}. Eye-tracking capabilities will likely be standard in future HMDs and, as such, Blinking is important to consider now. 

In our implementation, the user blinks both eyes at the beginning of the first letter to start the selection and then blinks again at the end of the last letter to complete the selection. Blinking of the two eyes (instead of the left/right eye) is used based on findings in \cite{hands_free_text_input}, whose text entry experiment shows that the accuracy of using two eyes is near 100\% (compared to 79\% with the left eye and 69\% with the right eye). Also, the literature suggests that a natural blink could last from 100-400ms depending on the situation. In VR, the blinking frequency is even higher \cite{eyeblink}. Therefore, we set 400ms as a minimal threshold to filter out natural and unintentional blinks. Our pilot tests suggest this threshold works well.

\subsubsection{Voice}
Voice activation has been used as an input modality in various user interfaces. There are two types of voice-based interactions: (1) voice command \cite{Voice_input}, and (2) nonverbal commands, like humming \cite{hummer} which rely on sound volume changes. Voice commands can be powerful but also come with many drawbacks, for instance, (1) voice recognition takes time (i.e., longer than nonverbal command recognition) and (2) people with speech disorders or with certain cognitive impairments cannot use it properly \cite{10.1145/1054972.1055013}. On the other hand, nonverbal commands that use sound volume can be easily detected and is much faster to process. Therefore, in this study, we focus on exploring nonverbal commands. We have followed \cite{hummer} to implement HumHum (a short hum at the first and the last letter of the text fragment) and Hummer (continuous humming from the first to the last letter of the text fragment) for text selection, with a volume threshold of 60db. We decided to only include Hummer in our experiment as our pilot results, in line with \cite{hummer}, showed that Hummer outperformed HumHum in performance and user experience. 

\subsubsection{Forward/Backward Neck Motions} 
Cursor movements can be controlled using lateral head motions or rotations. Forward and backward movements (or the depth dimension) could in theory also be used for activation or selection \cite{hands_free_text_input}. Yu et al. \cite{DepthMove} proposed DepthMove for VR which utilizes head forward and backward movements to indicate selection. They found that DepthMove is faster than dwelling (with a 1 dwell time) to select objects in 3D environments. This approach has also been explored in adaptive interfaces to adjust the level of detail given to users \cite{1383086}, to calibrate proximity-aware interactions \cite{10.1145/1357054.1357135}, and to enter text \cite{lu.2019.depthtext,hands_free_text_input}. In this study, we also implemented this method, followed the implementation in \cite{DepthMove} and \cite{hands_free_text_input}, and tested its performance and usability in pilot trials. Our pilot results show that these types of head motions are not suitable for text selection in VR because these types of motions are not precise enough when dealing with text content and, more importantly, participants dislike them as they found them tiring and uncomfortable to perform. Similar to findings from \cite{glanceable_feiyu}, participants found the motions not so acceptable to perform in public. Therefore, we did not include this in our experiment.

\begin{table}
\centering
  \caption{Rating for each usability criteria regarding the four selection mechanisms identified from the literature (0-3 ticks indicate ratings from worst to the best). UC1: Simple, easy, and fast to use; UC2: Minimal error rate and workload; UC3: Social acceptability.}
  \label{tab:freq}
  \begin{tabular}{p{2cm}p{1.5cm}p{1.5cm}p{1.5cm}}
    \toprule
    Mechanisms& UC1 & UC2 & UC3\\
    \midrule
    Dwell & \checkmark \checkmark \checkmark & \checkmark \checkmark \checkmark & \checkmark \checkmark \checkmark \\
    Eye blinks & \checkmark \checkmark \checkmark & \checkmark \checkmark \checkmark & \checkmark \checkmark \checkmark \\
    Voice & \checkmark \checkmark & \checkmark \checkmark \checkmark & \checkmark \\
    Neck & - & - & - \\

\bottomrule
\end{tabular}
\label{table:UC}
\end{table}

\subsection{Usability Criteria} \label{Section:UC}
In this section, we describe the following three usability criteria (UC) to determine the usefulness and suitability of each selection mechanism and, hence, to help narrow the scope of possible head-based text selection mechanisms for VR HMDs.

\begin{itemize}
    \item \textbf{\textit{UC1: Simple, easy, and fast to use}}. Selection usually goes together or complements pointer movement. With users controlling the pointer with their heads and paying attention to the highlighted text, selection should simple and straightforward so that it is possible to perform the two tasks almost at the same time. These two-step processes should be fast.
    \item \textbf{\textit{UC2: Minimal error rate and workload}}. Head-based interaction should be accurate, comfortable and convenient to perform \cite{pinpointing,6970627}, it is ideal for the selection mechanism to minimize the workload during text selection and have some degree of precision while allowing fast interaction. 
    
    \item \textbf{\textit{UC3: Social acceptability}}. While HMDs are designed for individual users, their interactions are often quite noticeable. As prior research has shown \cite{Fouad.2018.PerformerSocial, Pandey.2021.Speech, Vergari.2021.SocialEnv}, the social acceptability of interactions between HMDs and mobile devices can be an important factor in determining their adoption and usability.  
\end{itemize}

The rating process of the usability criteria of the four selection mechanisms (i.e., Dwell, Eye Blink, Voice, Neck; see Section \autoref{section:SelectionMethods}) was guided by findings reported in the literature and a pilot study. We used a 4-point Likert scale (with 0 indicating the worst and 3 the best). A summary of the rating for each selection mechanism can be found in \autoref{table:UC}. From the analysis of prior work together with the pilot study results, neck motions do not meet any of the criteria and were not considered further in the experiment. Other than the neck approach, Dwell, Eye blink, and Voice all meet at least two UC and hence are considered further and integrated into the experiment stage. This table will be revisited based on the results of our user study (see \autoref{table:UC:Results} in the Discussion section below). 


\section{Testbed Environment}
\begin{figure}[t]
  \centering
  \includegraphics[width=1\linewidth]{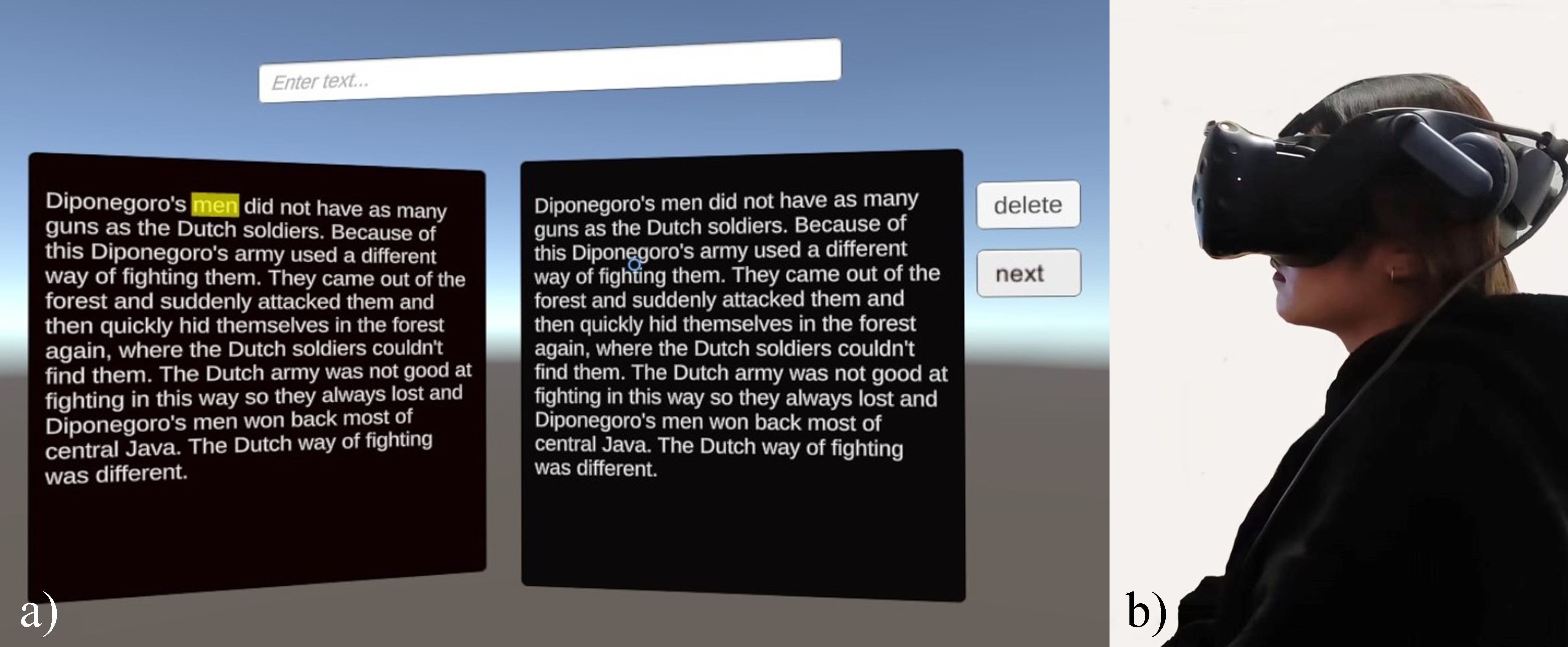}
  \caption{(a) A screenshot of the interface of the experiment environment which was used for all conditions. An ‘instruction panel’ is located on the left side, slightly tilted towards the user. The ‘interaction panel’ is located in the center, slightly tilted towards the user. Its design is based on recommendations from prior related work. (b) A picture of a participant doing the experiment with the HTC VIVE Pro Eye headset.}
  \label{fig:Testbed}
\end{figure}

Figure \ref{fig:Testbed} shows the test environment, which was also developed in Unity (v2019.4.10f1). An ‘instruction panel’ is located on the left side, where participants can see the text that needs to be selected. The ‘interaction panel’ is located at the center, where participants need to use each technique to select text fragments. In addition, two buttons are provided: ‘Delete’ for deleting the wrong selection, and ‘Next’ for moving to the next trial. 

The following parameters are set based on the recommendations from previous studies and then further tested and agreed upon by 5 users from a pilot study. We controlled the length of the materials to be between 9-12 lines, with each line having around 40 characters with spaces \cite{wei_reading_2020}, in both panels. The plane is set at 2.6m which our pilot participants have found to be suitable. 2.6m is also the recommended reading distance by \cite{dingler_vr_2018}. For the text style, we used Sans-serif Arial with a light color \cite{dingler_vr_2018}. Angular size was set as 1.8$^{\circ}$, which was within the recommended range suggested by \cite{dingler_vr_2018}. 

Visual support and feedback is provided in four ways: (1) The end of the ray is akin to a cursor, (2) changing the color of the selected text to yellow, (3) changing the color of the cursor when a selection was started/stopped, and (4) a visual indicator is provided for showing dwell progress. We did not display the ray because users from our pilot studies suggest the cursor alone is more effective than a combination of the ray and the cursor in helping them understand where they are pointing. In addition, they believe a head ray makes them feel overwhelmed because there are many visual changes on the display.

\section{Experiment}
\subsection{Participants and Apparatus}
24 participants (12 males, 12 females; aged 19-26) with a mean age of 21.75 (SD=1.66) from a local university campus volunteered to participate in this experiment. They all had normal or corrected-to-normal vision and did not have any difficulties moving their head or had any health issues that could affect their participation in the project. 21 of them have experienced VR HMDs, but only 9 were regular VR users (weekly) and 4 of them had interacted with the device used in this experiment, but they were not regular users of the HTC VIVE Pro Eye. 

The experimental application was run on a computer with an i7 processor, 16GB RAM, and an NVIDIA GTX 2080 Ti graphics card. An HTC VIVE Pro Eye VR headset was used in the experiment, which has a resolution of 2880 $\times$ 1600 pixels, 90 Hz refresh rate, and 110° (diagonal) FOV. The built-in Tobii eye-tracker was used to detect eye blinks with data transmission at 120Hz. The experiment was conducted in a quiet office room (30db). Participants were accompanied by one researcher and sat on a comfortable office chair during the experiment. 

\subsection{Design and Tasks}
The experiment followed a one-way within-subjects design with Interaction Techniques (Dwell, Blink, Sound; see \autoref{section:SelectionMethods} for their implementation details) as the independent variable.

\begin{figure}[t]
  \centering
  \includegraphics[width=1\linewidth]{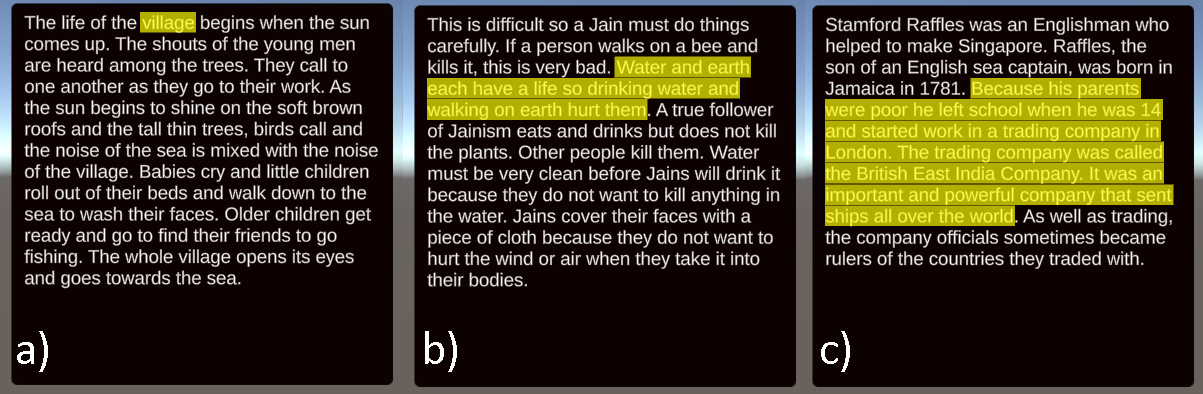}
  \caption{Screenshots of an example of short (a), medium (b), and long (c) text fragments used in our experiment.}
  \label{fig:length}
\end{figure}

For each condition, participants needed to complete 3 training trials (1 short, 1 medium, and 1 long text fragment; see Figure \ref{fig:length} for examples of each) and 27 trials (9 short, 9 medium, and 9 long texts) which were randomly sampled from a corpus of standardized English reading assessment \cite{quinn_asian_2007}. Each selection target would only appear once in a specific condition. The order of the \textit{interaction techniques} was counterbalanced across participants to avoid learning effects. Excluding the training texts, we collected 1944 trials (24 participants $\times$ 3 interaction techniques $\times$ 27 texts). Although the local area had nearly no local COVID-19 cases for 12 months before the experiment, we sanitized the device before and after each participant's turn and followed extra safety measures to ensure the safety of the participants and researchers (e.g., wearing a mask and staying at a safe distance and has good ventilation). 

\subsection{Procedure}
Participants were briefed on the goal of the research and the experimental procedure before the experiment began. Then, they needed to sign the consent form to participate in the experiment and filled out a demographic questionnaire (e.g., age, gender, and experience with VR). Before each condition started, the corresponding text selection method was explained to the participants, who then had a practice session with three warm-up selection tasks before the experiment stage (with 27 text selection tasks). Error correction was allowed by using the delete button in the VR scene (see Fig.~\ref{fig:Testbed}). The order of the conditions was balanced across participants. After each condition, participants needed to fill out a post-condition questionnaire (NASA-TLX \cite{hart_development_1988} and UEQ \cite{holzinger_construction_2008}). Once they completed the experiment, they needed to complete a post-experiment questionnaire and a structured interview. The whole experiment lasted around 30-40 minutes for each participant.

\subsection{Measurements}
We collected the following measurements to assess participants' performance and experience:

\begin{itemize}
\item \textit{Objective}: (1) \textit{Task-completion time}: The task completion time for each trial is defined as the time from when the cursor first hovers over the first target letter to the time they complete the correct selection. As such, the time spent on blinking, increasing the sound volume, and dwelling would be included for analysis. (2) \textit{Total error rate}: (the number of wrong sentences at the end + the number of deletions)/total number of attempts, (3) \textit{not corrected error rate}: the number of wrong sentences at the end/total number of sentences. 

Total error rate and not corrected error rate are measurement concepts derived from text entry studies \cite{10.1145/642611.642632}. The two measurements allow us to calculate the mistakes made during the text selection process in the data analysis (as participants are allowed to delete their selections if needed), instead of just counting how many mistakes were observed in the final recorded data. These two measurements provide a more complete picture of how many errors were made by the participants during the text selection tasks (deletions plus the number of error selections at the end) similar to text entry activities. Both completion time and error rates would help assess if and how well an approach meets \textit{UC1} and the first part of \textit{UC2} (see \autoref{Section:UC}).

\item \textit{Subjective}: \textit{NASA-TLX Questionnaire} \cite{hart_development_1988} to measure workload, \textit{User Experience Questionnaire} (UEQ) \cite{holzinger_construction_2008} to measure user experience, and collect participants' comments on the advantages and disadvantages of each technique plus their ranking of these. We used NASA-TLX and UEQ as they are commonly used to gather feedback from users in VR research \cite{Ringtext,Marco_selection_text,Pizzatext}. The NASA-TLX, UEQ, and participants' comments would allow determining how well each approach meets the second part of \textit{UC2} and \textit{UC3} (see \autoref{Section:UC}).
\end{itemize}

\subsection{Results}
Shapiro-Wilks tests and Q-Q plots were used to check if the data had a normal distribution. 

\textit{Performance analysis}. For normally distributed data, we employed two-way repeated-measures ANOVAs with \textit{Interaction Techniques} (Dwell, Blink, Voice) and \textit{Sentence Lengths} (short, medium, long) as the within-subjects variables. For data that were not normally distributed, we processed the data through Aligned Rank Transform (ART) \cite{Jacob_ART} before using repeated-measures ANOVAs with the transformed data. 

\textit{Experience analysis}. For normally distributed data, we employed one-way repeated-measures ANOVAs with \textit{Interaction Techniques} as the within-subjects variable. For data that were not normally distributed, like the performance analysis, we first processed the data through ART and then use repeated measure ANOVAs with the transformed data. 

For both analyses, we used Bonferroni correction for pairwise comparisons and Greenhouse-Geisser adjustment for degrees of freedom if there were violations of sphericity. All tests were with two-tailed p-values.

\subsubsection{Performance}
\textit{Task completion time}.
In total, we collected 1944 trials (24 participants $\times$ 3 interaction techniques $\times$ 27 texts) besides the training trials. To analyze task completion time, we discarded trials in which participants made a wrong selection (320 error trials or 16.4$\%$), and removed outliers, which were those trials whose selection time was more than three standard deviations from the mean (mean $\pm$ 3std.) in each condition (18 trials or 1.0$\%$). 

Table \ref{table:performance} shows the task completion time for each technique, where Blink is the fastest and Voice is the slowest technique. There was a statistically significant difference between Techniques on completion time ($F_{2,184}=51.427, p<.001$). Post-hoc analysis with Bonferroni correction suggested that Blink outperformed Dwell ($p<.001$) and Voice ($p<.001$), while Dwell outperformed Voice ($p<.001$).

We also observed a significant difference between Sentence Lengths on completion time ($F_{2,184}=10.831, p<.001$). Post-hoc analysis with Bonferroni correction suggested that participants completed Small faster than Medium ($p=.005$) and Long ($p<.0001$).

\textit{Total Error Rate (TER)}. 
Among these three techniques, Dwell achieved the best results (M=7.00$\%$, SD=6.69$\%$) and Voice had the worst results (M=11.40$\%$, SD=9.39$\%$). Table \ref{table:performance} shows more detailed results of TER. There was a statistically significant difference between Techniques on TER ($F_{2,184}=8.413, p<.001$). Post-hoc analysis with Bonferroni correction suggested that Voice was significantly worse than Blink ($p=.0015$) and Dwell ($p=.0014$). We did not observe any significant difference between Lengths ($F_{2,184}=0.367, p=.693$) nor the interaction of Techniques $\times$ Lengths ($F_{4,184}=0.554, p=.554$).

\textit{Not Corrected Error Rate (NCER)} 
In general, among these three techniques, Voice achieved the lowest NCER (M=0.67 $\%$, SD=1.98$\%$) and Blink had the highest NCER (M=1.17 $\%$, SD=2.68$\%$); see Table \ref{table:performance} for more details. There was a statistically significant difference between Techniques on NCER ($F_{2,184}=6.935, p=.0012$). Post-hoc analysis with Bonferroni correction suggested that Blink was worse than Dwell ($p=.0048$) and Voice ($p=.0041$). We could not find any significant effect of Lengths ($F_{2,184}=0.743, p=.477$) on NCER and the interaction of Techniques $\times$ Lengths ($F_{4,184}=0.344, p=.848$) on NCER.

\begin{table*}
\centering
  \caption{Performance data for each Interaction Technique among three Sentence Lengths, mean (SD). The ranking of each condition is indicated with Roman numerals (I: \colorbox{mycolor1}{light green}; II: \colorbox{mycolor2}{darker light green}; and III: \colorbox{mycolor3}{blue-green}).}
  \label{tab:freq}
  \begin{tabular}{p{3.5cm} p{2.5cm} p{2.2cm}p{2.2cm}p{2.2cm}}
    \toprule
    Performance Metrics&Sentence Length& Blink & Dwell & Voice\\
    \midrule
    Task completion time & Small& \cellcolor{mycolor1}I: 2.32 (0.50)& \cellcolor{mycolor2}II: 2.51 (0.36)& \cellcolor{mycolor3}III: 3.16 (0.94)\\
      & Medium & \cellcolor{mycolor1}I: 2.32 (0.57) & \cellcolor{mycolor2}II: 3.03 (0.54)& \cellcolor{mycolor3}III: 3.52 (0.90)\\
      & Long & \cellcolor{mycolor1}I: 2.60(0.69)& \cellcolor{mycolor2}II: 3.10 (0.46)& \cellcolor{mycolor3}III: 3.54 (1.12)\\
    TER  & Small & \cellcolor{mycolor1}I: 7.4\% (9.3\%) & \cellcolor{mycolor2}II: 7.9\% (7.1\%) & \cellcolor{mycolor3}III: 11.6\% (9.6\%) \\
         & Medium & \cellcolor{mycolor1}I: 6.9\% (6.4\%)& \cellcolor{mycolor2}II: 7.3\% (6.4\%)& \cellcolor{mycolor3}III: 12.1\% (9.8\%)\\
      & Long & \cellcolor{mycolor2}II: 8.8\% (9.5\%) & \cellcolor{mycolor1}I: 5.8\% (6.4\%) & \cellcolor{mycolor3}III: 10.5\% (8.7\%) \\
   NCER  & Small & \cellcolor{mycolor3}III: 1.2\% (2.5\%) & \cellcolor{mycolor2}II: 1.1\% (1.7\%) & \cellcolor{mycolor1}I: 0.8\% (2.4\%) \\
         & Medium & \cellcolor{mycolor3}III: 1.2\% (2.8\%) & \cellcolor{mycolor1}I: 0.9\% (1.9\%) & \cellcolor{mycolor2}II: 1.1\% (2.3\%) \\
      & Long & \cellcolor{mycolor3}III: 1.1\% (2.7\%) & \cellcolor{mycolor2}II: 0.3\% (1.0\%) & \cellcolor{mycolor1}I: 0.1\% (0.7\%) \\
\bottomrule
\end{tabular}
\label{table:performance}
\end{table*}

\subsubsection{User Experience}
\textit{UEQ}. For average scores, Blink achieved the best results (M=1.51, SD=0.23), Voice was the second (M=1.18, SD=0.39), and Dwell had the worst results (M=1.14, SD=0.36). However, ANOVA tests showed no significant difference between Techniques ($F_{2, 46}=2.228, p=0.119$). Regarding the UEQ subscales, ANOVA tests yielded a significant difference between Techniques on Attractiveness ($F_{2,46}=4.093, p=.023$), Efficiency ($F_{2,46}=3.837, p=.029$), and Novelty ($F_{2,46}=4.192, p=.021$). Post-hoc pairwise comparison suggested that Voice had a significantly higher score than Dwell in Novelty ($p=.030$). However, post-hoc tests did not yield any difference between Techniques regarding Attractiveness and Efficiency. We could not find any significant effect of Techniques on Perspicuity ($F_{1.445,33.242}=.953, p=.369$), Dependability ($F_{2,46}=1.189, p=.314$), and Stimulation ($F_{2,46}=.520, p=.598$). 

\begin{figure}[t]
  \centering
  \includegraphics[width=1\linewidth]{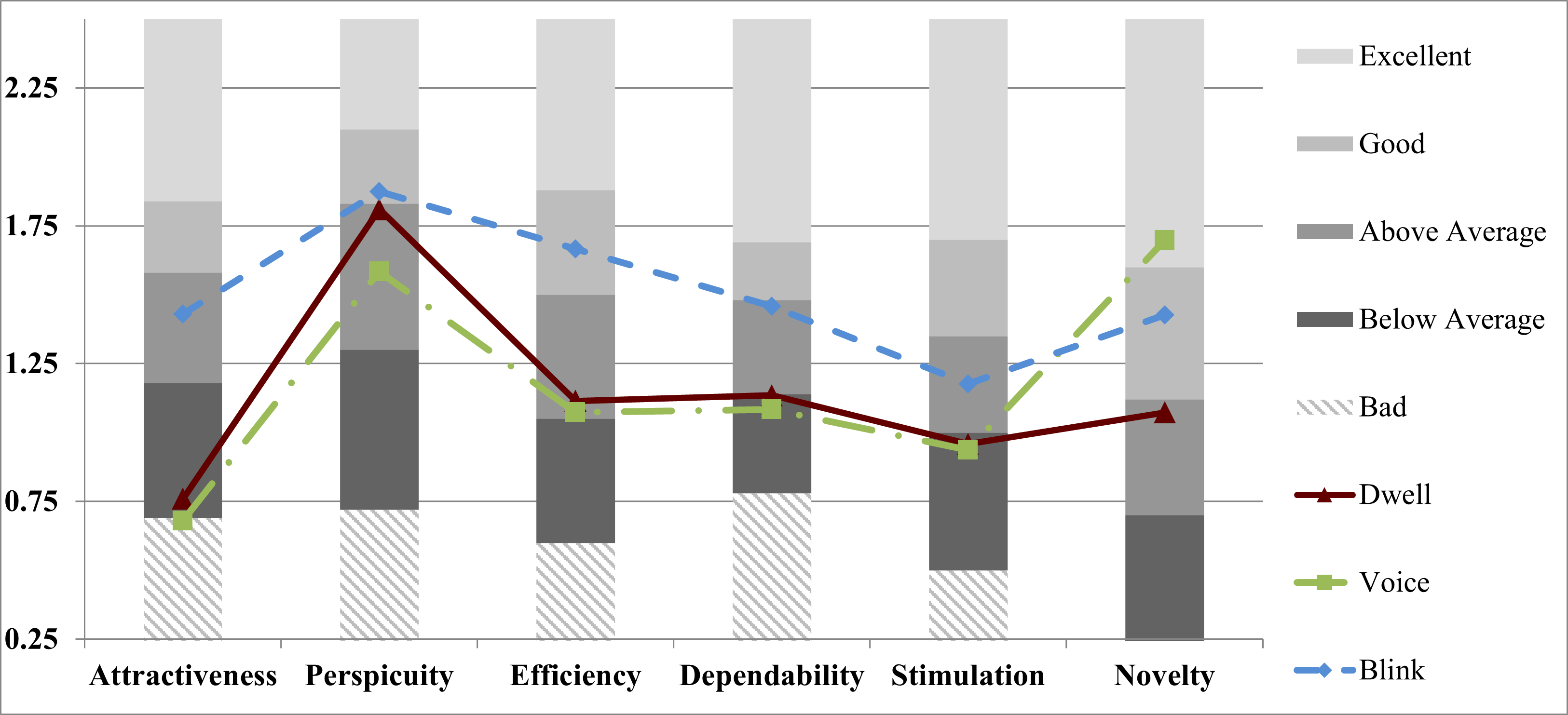}
  \caption{UEQ subscale ratings of the tested methods concerning comparison benchmarks.}
  \label{fig:UEQbenchmark}
\end{figure}

Details of each UEQ subscale score can be found in Figure \ref{fig:UEQbenchmark}. Overall Blink was rated Above Average to Good; Dwell was rated Below Average to Above Average, and Voice was rated from Bad to Excellent (mainly Below Average to Above Average).

\textit{Workload}. For overall task workload, Blink was rated the best (M=21.46, SD=16.10), Dwell the second (M=30.47, SD=19.05), and Voice the worst (M=35.45, SD=20.72). ANOVA tests yielded a significant difference between Techniques ($F_{2,46}=6.173, p=.004$). Post-hoc pairwise comparisons indicated that users experienced less overall workload with Blink than with Voice ($p=.013$). 

\begin{figure}[t]
  \centering
  \includegraphics[width=1\linewidth]{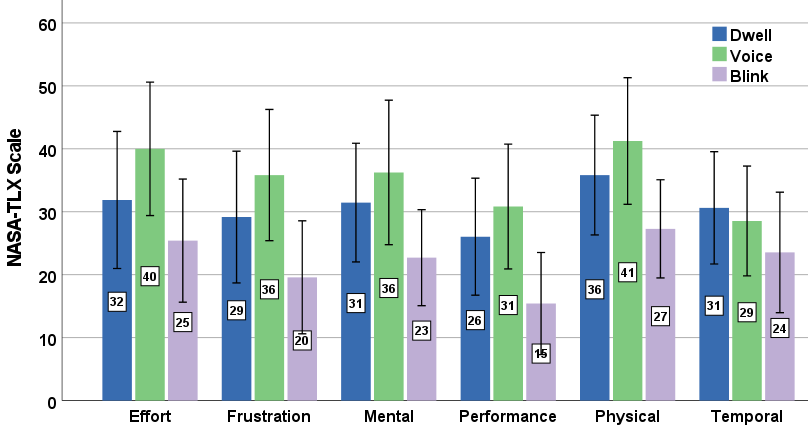}
  \caption{The mean responses for the 6 components of the NASA TLX questionnaire. Error bars indicate a 95\% confidence interval.}
  \label{fig:NASA}
\end{figure}

Regarding each NASA-TLX workload subscale, ANOVA tests yielded significant effects between Techniques on Physical ($F_{2,46}=4.535, p=.016$), Performance ($F_{2,46}=5.675, p=.006$), Mental ($F_{2,46}=3.978, p=.026$) and Frustration ($F_{2,46}=3.847, p=.029$). Post-hoc pairwise comparisons suggested that participants believed Blink led to less workload than Voice regarding (1) Physical ($p=.040$), (2) Performance ($p=.016$), and (3) Mental ($p=.034$). Post-hoc analysis yielded no significant difference between Techniques regarding Frustration. Details of NASA-TLX workload subscales can be found in Fig. \ref{fig:NASA}.

\textit{Ranking}. Users' rankings show a preference for Blink (17 ranked it first while 6 ranked it second) among the three techniques. It was followed by Dwell (7 ranked it first while 11 ranked it second). Voice was generally rated the worst (8 ranked it second and 16 ranked it third). 

\subsubsection{Qualitative Feedback} In general, most participants stated positive comments about Blink: "fast" (N=8), "simple" (N=3), "convenient" (N=7), and "accurate" (N=6). There was one negative comment: "frequent blinking caused eye discomfort" (N=1). Similarly, participants comment positively about Dwell: "convenient" (N=5). Some commented that it was "hard/difficult to stay still" during selection (N=4). In addition, some participants (N=5) described Voice as "creative", but more said it as "embarrassing" and as such, they were not so willing to use it in "public places" (N=7) due to the need to constantly make noises. In general, participants said that both Blink and Dwell were socially acceptable (that is, they would use them in front of others). On the other hand, they would use Voice in private places without anyone around them.

\section{Discussion}
In general, a dwell-based approach is a typical, default choice for hands-free interaction because it is relatively easy to use and does not require additional hardware. However, dwell has inherent issues, for example, if dwell time is long, it makes interaction inefficient and if it is short, it can lead to errors \cite{dwell_tradeoff, hands_free_text_input}---in short, users have less control over the process. With the rapid advances of HMDs, new sensing capabilities have been added to these, particularly eye-tracking, which should be more and more standard. In this work, we focused on comparing the dwell-based selection mechanism by using eye blinks and voice. Both of them are relatively easy to do, give more control to users, and, more importantly, can be captured by a wide range of HMDs without additional hardware requirements or external sensing (handheld) devices, unlike some recent work \cite{darbar_exploring_2021, lee_one-thumb_2020, Lee.2021CopyPaste}, which can add extra costs to users and complexity to the interaction process.

Overall, our results suggest that Blink has the best performance (speed and accuracy) in text selection tasks and outperforms Dwell and Voice, which aligns with \cite{hands_free_text_input, Lu.2021.Itext} for hands-free text entry tasks in both AR and VR, where Blink outperforms Dwell. Voice has the worst performance in task completion time and total error rate. Based on our observations, we believe the reasons could be due to (1) users' unstable voice volume control: Keeping the voice volume at/above 60db during the selection process can be difficult for participants, which has led to an increase in errors, (2) users need to inhale: users are only able to exhale when making a sound using the Hummer technique; if users are not well trained for and familiar with this technique, they have to stop in the middle of the selection process to inhale oxygen. Regarding the effect of sentence length on performance, we found that the length of a sentence has a significant effect on the completion time, as longer sentences require more time to complete their selection. 

In line with \cite{hands_free_text_input}, we could not find any significant difference between Blink and Dwell regarding user experience. In general, Blink has the lowest workload ratings and outperforms Voice (overall workload, physical, performance, and mental). We could not find a significant difference between Blink and Dwell, suggesting that intentional eye blinks might be as acceptable as Dwell regarding workload. In summary, Blink should be given priority for the lowest workload.

In \autoref{Section:UC} we described three usability criteria (UC) that we have identified from the literature. Based on the above results, we revisited \autoref{table:UC}. Of the three mechanisms, Blink is the only one that has kept the same ratings unchanged. Dwell received a slightly lower rating for UC1 and UC2. Voice has been found to be not so acceptable in social settings, to be difficult to use, and lead to higher errors and workload.

\begin{table}
\centering
  \caption{Rating for each usability criteria regarding the three selection mechanisms tested in our experiment (0-3 ticks indicate ratings from worst to the best). UC1: Simple, easy, and fast to use; UC2: Minimal error rate and workload; UC3: Social acceptability.}
  \label{tab:freq}
  \begin{tabular}{p{2cm}p{1.5cm}p{1.5cm}p{1.5cm}}
    \toprule
    Mechanisms& UC1 & UC2 & UC3\\
    \midrule
    Dwell & \checkmark \checkmark & \checkmark \checkmark & \checkmark \checkmark \checkmark \\
    Eye blinks & \checkmark \checkmark \checkmark & \checkmark \checkmark \checkmark & \checkmark \checkmark \checkmark \\
    Voice & \checkmark & \checkmark & - \\
\bottomrule
\end{tabular}
\label{table:UC:Results}
\end{table}

\subsection{Key Takeaways and Lessons}

\begin{itemize}
\item \textit{Blink} is a viable solution for hands-free text selection in VR/AR HMDs; it could possibly be useful for text editing in general as it has also been found to work well for text entry \cite{hands_free_text_input, Lu.2021.Itext}. It could be set as default if an eye tracker is available because it has the best performance, good user experience, and low workload/error rate, and is acceptable in public settings.
\item \textit{Dwell} is an acceptable hands-free method and should be used when an eye-tracker is unavailable or when the HMD does not have eye-tracking capabilities (which is still a common scenario for low-cost but popular VR HMDs such as Meta Quest 2).
\item \textit{Voice input} such as nonverbal volume-based input should be avoided due to the poor performance and its perceived socially unacceptability as mentioned by participants. 
\end{itemize}

\subsection{Limitations and Future Work}
There are some limitations of this study. Due to a lack of standard performance metrics for this type of study, we measured the speed through optimal task completion time from pointing selection tasks \cite{pointingstudy1,pointingstudy2} and measured error rates using the concepts of total error rate and not corrected error rate derived from text entry studies \cite{10.1145/642611.642632,xu_pointing_2019}. Given the limited work on performance metrics for text selection, future work is needed to establish standard metrics for text selection experiments. 

In addition, some design considerations are derived from prior related work (e.g., dwell duration, white text highlighted in yellow, display of the ray, interaction distance) and determined through with pilot study users. Future exploration of these factors would be useful for the development of new techniques that can improve user performance further and increase their acceptability of the techniques. We only included university students as our participants in this study and future work can use a more inclusive and diverse population (e.g., impaired and elderly users who might have difficulty using their hands). Future work can also involve longitudinal explorations of text selection approaches that consist of more sessions (e.g., 1 or 2 sessions for 4-5 days like some studies that explored text entry in VR/AR \cite{Ringtext,Pizzatext,wristext,Lu.2021.Itext}). Also, as eye trackers become a standard feature in many HMDs (e.g., Pico Neo, HoloLens 2, Magic Leap), it would be useful to explore eye gaze in the future and compare it against head-based pointing.

The movement toward an office that is more virtual and immersive \cite{grubert_office_2018} is gaining rapid momentum with the development of the Metaverse, where users need to use common office applications that depend heavily on text editing (e.g., in spreadsheets \cite{Grubert.2020.Spreesheets} and presentations \cite{Grubert2022.PoVRPoint}). However, text editing in VR/AR, of which text selection is an important aspect, is far behind in the level of efficiency and usability to which people are used in less immersive, traditional platforms, like desktop and laptop computers. In the future, we plan to extend our work on text selection, still focusing on hands-free approaches as this work shows they are usable and efficient, and integrate the results into the ecosystem of text editing, especially in combination with text entry techniques that are also hands-free (e.g., \cite{hands_free_text_input, Lu.2021.Itext, Ringtext}).   

\section{Conclusion}
In this work, we have implemented three selection mechanisms (Voice, Blink, Dwell) derived from our review of the literature on hands-free interaction for virtual reality (VR) head-mounted displays (HMDs) for text selection tasks. Our results with 24 participants showed that an approach based on eye blinks is the best hands-free selection candidate as it has the fastest performance, best accuracy, highest experience, lowest workload, and positive social acceptability. It is followed by a dwell-like approach, which has better accuracy than using voice. According to our results, eye blinks can be an excellent hands-free selection mechanism and should be used for text selection in VR if an eye tracker is available.

\acknowledgments{
The authors want to thank the participants who joined the study and the reviewers for their insightful comments and useful suggestions that helped improve our paper. This work was supported in part by Xi'an Jiaotong-Liverpool University (XJTLU) Key Special Fund (\#KSF-A-03) and XJTLU Research Development Fund (\#RDF-17-01-54).}

\bibliographystyle{abbrv-doi}

\bibliography{template}

\begin{thebibliography}{10}

\bibitem{Fouad.2018.PerformerSocial}
F.~Alallah, A.~Neshati, Y.~Sakamoto, K.~Hasan, E.~Lank, A.~Bunt, and P.~Irani.
\newblock Performer vs. observer: Whose comfort level should we consider when
  examining the social acceptability of input modalities for head-worn display?
\newblock In {\em Proceedings of the 24th ACM Symposium on Virtual Reality
  Software and Technology}, VRST '18, p.~9. Association for Computing
  Machinery, New York, NY, USA, 2018. doi: {{%
10\hspace{.1pt}\discretionary{.}{%
}{.}\hspace{.4pt}1145\discretionary{/}{%
}{/}3281505\hspace{.1pt}\discretionary{.}{%
}{.}\hspace{.4pt}3281541}}


\bibitem{baird_evaluating_1999}
K.~M. Baird and W.~Barfield.
\newblock Evaluating the effectiveness of augmented reality displays for a
  manual assembly task.
\newblock {\em Virtual Reality}, 4(4):10, Dec. 1999. doi: {{%
10\hspace{.1pt}\discretionary{.}{%
}{.}\hspace{.4pt}1007\discretionary{/}{%
}{/}BF01421808}}


\bibitem{Grubert2022.PoVRPoint}
V.~Biener, T.~Gesslein, D.~Schneider, F.~Kawala, A.~Otte, P.~O. Kristensson,
  M.~Pahud, E.~Ofek, C.~Campos, M.~Kljun, K.~Ä. Pucihar, and J.~Grubert.
\newblock Povrpoint: Authoring presentations in mobile virtual reality.
\newblock {\em IEEE Transactions on Visualization and Computer Graphics},
  28(5):2069--2079, 2022. doi: {{%
10\hspace{.1pt}\discretionary{.}{%
}{.}\hspace{.4pt}1109\discretionary{/}{%
}{/}TVCG\hspace{.1pt}\discretionary{.}{%
}{.}\hspace{.4pt}2022\hspace{.1pt}\discretionary{.}{%
}{.}\hspace{.4pt}3150474}}


\bibitem{183317}
T.~Caudell and D.~Mizell.
\newblock Augmented reality: an application of heads-up display technology to
  manual manufacturing processes.
\newblock In {\em Proceedings of the Twenty-Fifth Hawaii International
  Conference on System Sciences}, vol.~ii, p.~11, 1992. doi: {{%
10\hspace{.1pt}\discretionary{.}{%
}{.}\hspace{.4pt}1109\discretionary{/}{%
}{/}HICSS\hspace{.1pt}\discretionary{.}{%
}{.}\hspace{.4pt}1992\hspace{.1pt}\discretionary{.}{%
}{.}\hspace{.4pt}183317}}


\bibitem{blink_selection}
I.~Chatterjee, R.~Xiao, and C.~Harrison.
\newblock Gaze+gesture: Expressive, precise and targeted free-space
  interactions.
\newblock In {\em Proceedings of the 2015 ACM on International Conference on
  Multimodal Interaction}, ICMI '15, p.~8. Association for Computing Machinery,
  New York, NY, USA, 2015. doi: {{%
10\hspace{.1pt}\discretionary{.}{%
}{.}\hspace{.4pt}1145\discretionary{/}{%
}{/}2818346\hspace{.1pt}\discretionary{.}{%
}{.}\hspace{.4pt}2820752}}


\bibitem{lei.2021.pointing}
L.~Chen, Y.~Liu, Y.~Li, L.~Yu, B.~Gao, M.~Caon, Y.~Yue, and H.-N. Liang.
\newblock Effect of visual cues on pointing tasks in co-located augmented
  reality collaboration.
\newblock In {\em Symposium on Spatial User Interaction}, SUI '21, p.~12.
  Association for Computing Machinery, New York, NY, USA, 2021. doi: {{%
10\hspace{.1pt}\discretionary{.}{%
}{.}\hspace{.4pt}1145\discretionary{/}{%
}{/}3485279\hspace{.1pt}\discretionary{.}{%
}{.}\hspace{.4pt}3485297}}


\bibitem{darbar_exploring_2021}
R.~Darbar, A.~Prouzeau, J.~Odicio-Vilchez, T.~Lainé, and M.~Hachet.
\newblock Exploring {Smartphone}-enabled {Text} {Selection} in {AR}-{HMD}.
\newblock In {\em Proceedings of {Graphics} {Interface} 2021}, {GI} 2021, pp.
  117 -- 126. Canadian Information Processing Society, 2021.
\newblock ISSN: 0713-5424 event-place: Virtual Event. doi: {{%
10\hspace{.1pt}\discretionary{.}{%
}{.}\hspace{.4pt}20380\discretionary{/}{%
}{/}GI2021\hspace{.1pt}\discretionary{.}{%
}{.}\hspace{.4pt}14}}


\bibitem{dingler_vr_2018}
T.~Dingler, K.~Kunze, and B.~Outram.
\newblock {VR} {Reading} {UIs}: {Assessing} {Text} {Parameters} for {Reading}
  in {VR}.
\newblock In {\em Extended {Abstracts} of the 2018 {CHI} {Conference} on
  {Human} {Factors} in {Computing} {Systems}}, {CHI} {EA} '18, p.~6.
  Association for Computing Machinery, New York, NY, USA, 2018.
\newblock event-place: Montreal QC, Canada. doi: {{%
10\hspace{.1pt}\discretionary{.}{%
}{.}\hspace{.4pt}1145\discretionary{/}{%
}{/}3170427\hspace{.1pt}\discretionary{.}{%
}{.}\hspace{.4pt}3188695}}


\bibitem{1383086}
S.~DiVerdi, T.~Hollerer, and R.~Schreyer.
\newblock Level of detail interfaces.
\newblock In {\em Third IEEE and ACM International Symposium on Mixed and
  Augmented Reality}, p.~2, 2004. doi: {{%
10\hspace{.1pt}\discretionary{.}{%
}{.}\hspace{.4pt}1109\discretionary{/}{%
}{/}ISMAR\hspace{.1pt}\discretionary{.}{%
}{.}\hspace{.4pt}2004\hspace{.1pt}\discretionary{.}{%
}{.}\hspace{.4pt}38}}


\bibitem{ESTEVES2020102414}
A.~Esteves, Y.~Shin, and I.~Oakley.
\newblock Comparing selection mechanisms for gaze input techniques in
  head-mounted displays.
\newblock {\em International Journal of Human-Computer Studies}, 139:10, 2020.
  doi: {{%
10\hspace{.1pt}\discretionary{.}{%
}{.}\hspace{.4pt}1016\discretionary{/}{%
}{/}j\hspace{.1pt}\discretionary{.}{%
}{.}\hspace{.4pt}ijhcs\hspace{.1pt}\discretionary{.}{%
}{.}\hspace{.4pt}2020\hspace{.1pt}\discretionary{.}{%
}{.}\hspace{.4pt}102414}}


\bibitem{Grubert.2020.Spreesheets}
T.~Gesslein, V.~Biener, P.~Gagel, D.~Schneider, P.~O. Kristensson, E.~Ofek,
  M.~Pahud, and J.~Grubert.
\newblock Pen-based interaction with spreadsheets in mobile virtual reality.
\newblock In {\em 2020 IEEE International Symposium on Mixed and Augmented
  Reality (ISMAR)}, pp. 361--373, 2020. doi: {{%
10\hspace{.1pt}\discretionary{.}{%
}{.}\hspace{.4pt}1109\discretionary{/}{%
}{/}ISMAR50242\hspace{.1pt}\discretionary{.}{%
}{.}\hspace{.4pt}2020\hspace{.1pt}\discretionary{.}{%
}{.}\hspace{.4pt}00063}}


\bibitem{ghosh_eyeditor_2020}
D.~Ghosh, P.~S. Foong, S.~Zhao, C.~Liu, N.~Janaka, and V.~Erusu.
\newblock {EYEditor}: {Towards} {On}-the-{Go} {Heads}-{Up} {Text} {Editing}
  {Using} {Voice} and {Manual} {Input}.
\newblock In {\em Proceedings of the 2020 {CHI} {Conference} on {Human}
  {Factors} in {Computing} {Systems}}, pp. 1--13. Association for Computing
  Machinery, New York, NY, USA, 2020. doi: {{%
10\hspace{.1pt}\discretionary{.}{%
}{.}\hspace{.4pt}1145\discretionary{/}{%
}{/}3313831\hspace{.1pt}\discretionary{.}{%
}{.}\hspace{.4pt}3376173}}


\bibitem{wristext}
J.~Gong, Z.~Xu, Q.~Guo, T.~Seyed, X.~A. Chen, X.~Bi, and X.-D. Yang.
\newblock Wristext: One-handed text entry on smartwatch using wrist gestures.
\newblock In {\em Proceedings of the 2018 CHI Conference on Human Factors in
  Computing Systems}, CHI '18, p.~14. Association for Computing Machinery, New
  York, NY, USA, 2018. doi: {{%
10\hspace{.1pt}\discretionary{.}{%
}{.}\hspace{.4pt}1145\discretionary{/}{%
}{/}3173574\hspace{.1pt}\discretionary{.}{%
}{.}\hspace{.4pt}3173755}}


\bibitem{grauman_communication_2003}
K.~Grauman, M.~Betke, J.~Lombardi, J.~Gips, and G.~Bradski.
\newblock Communication via eye blinks and eyebrow raises: video-based
  human-computer interfaces.
\newblock {\em Universal Access in the Information Society}, 2(4):359--373,
  Nov. 2003. doi: {{%
10\hspace{.1pt}\discretionary{.}{%
}{.}\hspace{.4pt}1007\discretionary{/}{%
}{/}s10209\discretionary{%
}{-}{-}003\discretionary{%
}{-}{-}0062\discretionary{%
}{-}{-}x}}


\bibitem{grubert_office_2018}
J.~Grubert, E.~Ofek, M.~Pahud, and P.~O. Kristensson.
\newblock The {Office} of the {Future}: {Virtual}, {Portable}, and {Global}.
\newblock {\em IEEE Computer Graphics and Applications}, 38(6):125--133, Nov.
  2018. doi: {{%
10\hspace{.1pt}\discretionary{.}{%
}{.}\hspace{.4pt}1109\discretionary{/}{%
}{/}MCG\hspace{.1pt}\discretionary{.}{%
}{.}\hspace{.4pt}2018\hspace{.1pt}\discretionary{.}{%
}{.}\hspace{.4pt}2875609}}


\bibitem{grubert2018text}
J.~Grubert, L.~Wizani, E.~Ofek, M.~Pahud, M.~Kranz, and P.~O. Krisstensson.
\newblock Text entry in immersive head-mounted display-based virtual reality
  using physical and touch keyboards.
\newblock In {\em IEEE VR 2018}, p.~8. IEEE, March 2018.

\bibitem{eyeblink}
D.~Gurung, G.~Kc, and P.~ADHIKARY.
\newblock Mathematical model of thermal effects of blinking in human eye.
\newblock {\em International Journal of Biomathematics}, 9:20, 06 2015. doi:
  {{%
10\hspace{.1pt}\discretionary{.}{%
}{.}\hspace{.4pt}1142\discretionary{/}{%
}{/}S1793524516500066}}


\bibitem{10.1145/1357054.1357135}
C.~Harrison and A.~K. Dey.
\newblock Lean and zoom: Proximity-aware user interface and content
  magnification.
\newblock In {\em Proceedings of the SIGCHI Conference on Human Factors in
  Computing Systems}, CHI '08, p.~4. Association for Computing Machinery, New
  York, NY, USA, 2008. doi: {{%
10\hspace{.1pt}\discretionary{.}{%
}{.}\hspace{.4pt}1145\discretionary{/}{%
}{/}1357054\hspace{.1pt}\discretionary{.}{%
}{.}\hspace{.4pt}1357135}}


\bibitem{hart_development_1988}
S.~G. Hart and L.~E. Staveland.
\newblock Development of {NASA}-{TLX} ({Task} {Load} {Index}): {Results} of
  empirical and theoretical research.
\newblock In {\em Advances in psychology}, vol.~52, pp. 139--183. Elsevier,
  1988.

\bibitem{hummer}
R.~Hedeshy, C.~Kumar, R.~Menges, and S.~Staab.
\newblock Hummer: Text entry by gaze and hum.
\newblock In {\em Proceedings of the 2021 CHI Conference on Human Factors in
  Computing Systems}, CHI '21, p.~11. Association for Computing Machinery, New
  York, NY, USA, 2021. doi: {{%
10\hspace{.1pt}\discretionary{.}{%
}{.}\hspace{.4pt}1145\discretionary{/}{%
}{/}3411764\hspace{.1pt}\discretionary{.}{%
}{.}\hspace{.4pt}3445501}}


\bibitem{pointingstudy2}
J.~Huang, F.~Tian, X.~Fan, X.~L. Zhang, and S.~Zhai.
\newblock Understanding the uncertainty in 1d unidirectional moving target
  selection.
\newblock In {\em Proceedings of the 2018 CHI Conference on Human Factors in
  Computing Systems}, CHI '18, p. 1–12. Association for Computing Machinery,
  New York, NY, USA, 2018. doi: {{%
10\hspace{.1pt}\discretionary{.}{%
}{.}\hspace{.4pt}1145\discretionary{/}{%
}{/}3173574\hspace{.1pt}\discretionary{.}{%
}{.}\hspace{.4pt}3173811}}


\bibitem{Dwell_Technique}
R.~J.~K. Jacob.
\newblock What you look at is what you get: Eye movement-based interaction
  techniques.
\newblock In {\em Proceedings of the SIGCHI Conference on Human Factors in
  Computing Systems}, CHI '90, p.~8. Association for Computing Machinery, New
  York, NY, USA, 1990. doi: {{%
10\hspace{.1pt}\discretionary{.}{%
}{.}\hspace{.4pt}1145\discretionary{/}{%
}{/}97243\hspace{.1pt}\discretionary{.}{%
}{.}\hspace{.4pt}97246}}


\bibitem{dwell_tradeoff}
R.~J.~K. Jacob.
\newblock The use of eye movements in human-computer interaction techniques:
  What you look at is what you get.
\newblock {\em ACM Trans. Inf. Syst.}, 9(2):152–169, apr 1991. doi: {{%
10\hspace{.1pt}\discretionary{.}{%
}{.}\hspace{.4pt}1145\discretionary{/}{%
}{/}123078\hspace{.1pt}\discretionary{.}{%
}{.}\hspace{.4pt}128728}}


\bibitem{6970627}
S.~Jalaliniya, D.~Mardanbeigi, T.~Pederson, and D.~W. Hansen.
\newblock Head and eye movement as pointing modalities for eyewear computers.
\newblock In {\em 2014 11th International Conference on Wearable and
  Implantable Body Sensor Networks Workshops}, pp. 50--53, 2014. doi: {{%
10\hspace{.1pt}\discretionary{.}{%
}{.}\hspace{.4pt}1109\discretionary{/}{%
}{/}BSN\hspace{.1pt}\discretionary{.}{%
}{.}\hspace{.4pt}Workshops\hspace{.1pt}\discretionary{.}{%
}{.}\hspace{.4pt}2014\hspace{.1pt}\discretionary{.}{%
}{.}\hspace{.4pt}14}}


\bibitem{pinpointing}
M.~Kyt\"{o}, B.~Ens, T.~Piumsomboon, G.~A. Lee, and M.~Billinghurst.
\newblock Pinpointing: Precise head- and eye-based target selection for
  augmented reality.
\newblock In {\em Proceedings of the 2018 CHI Conference on Human Factors in
  Computing Systems}, CHI '18, p.~14. Association for Computing Machinery, New
  York, NY, USA, 2018. doi: {{%
10\hspace{.1pt}\discretionary{.}{%
}{.}\hspace{.4pt}1145\discretionary{/}{%
}{/}3173574\hspace{.1pt}\discretionary{.}{%
}{.}\hspace{.4pt}3173655}}


\bibitem{holzinger_construction_2008}
B.~Laugwitz, T.~Held, and M.~Schrepp.
\newblock Construction and {Evaluation} of a {User} {Experience}
  {Questionnaire}.
\newblock In A.~Holzinger, ed., {\em {HCI} and {Usability} for {Education} and
  {Work}}, vol. 5298, p.~14. Springer Berlin Heidelberg, Berlin, Heidelberg,
  2008. doi: {{%
10\hspace{.1pt}\discretionary{.}{%
}{.}\hspace{.4pt}1007\discretionary{/}{%
}{/}978\discretionary{%
}{-}{-}3\discretionary{%
}{-}{-}540\discretionary{%
}{-}{-}89350\discretionary{%
}{-}{-}9\_6}}


\bibitem{lee_one-thumb_2020}
L.-H. Lee, Y.~Zhu, Y.-P. Yau, T.~Braud, X.~Su, and P.~Hui.
\newblock One-thumb {Text} {Acquisition} on {Force}-assisted {Miniature}
  {Interfaces} for {Mobile} {Headsets}.
\newblock In {\em 2020 {IEEE} {International} {Conference} on {Pervasive}
  {Computing} and {Communications} ({PerCom})}, pp. 1--10, 2020. doi: {{%
10\hspace{.1pt}\discretionary{.}{%
}{.}\hspace{.4pt}1109\discretionary{/}{%
}{/}PerCom45495\hspace{.1pt}\discretionary{.}{%
}{.}\hspace{.4pt}2020\hspace{.1pt}\discretionary{.}{%
}{.}\hspace{.4pt}9127378}}


\bibitem{Lee.2021CopyPaste}
L.~H. Lee, Y.~Zhu, Y.-P. Yau, P.~Hui, and S.~Pirttikangas.
\newblock Press-n-paste: Copy-and-paste operations with pressure-sensitive
  caret navigation for miniaturized surface in mobile augmented reality.
\newblock {\em Proc. ACM Hum.-Comput. Interact.}, 5(EICS):29, may 2021. doi:
  {{%
10\hspace{.1pt}\discretionary{.}{%
}{.}\hspace{.4pt}1145\discretionary{/}{%
}{/}3457146}}


\bibitem{glanceable_feiyu}
F.~Lu, S.~Davari, and D.~Bowman.
\newblock Exploration of techniques for rapid activation of glanceable
  information in head-worn augmented reality.
\newblock In {\em Symposium on Spatial User Interaction}, SUI '21, p.~11.
  Association for Computing Machinery, New York, NY, USA, 2021. doi: {{%
10\hspace{.1pt}\discretionary{.}{%
}{.}\hspace{.4pt}1145\discretionary{/}{%
}{/}3485279\hspace{.1pt}\discretionary{.}{%
}{.}\hspace{.4pt}3485286}}


\bibitem{lu.2019.depthtext}
X.~Lu, D.~Yu, H.-N. Liang, X.~Feng, and W.~Xu.
\newblock Depthtext: Leveraging head movements towards the depth dimension for
  hands-free text entry in mobile virtual reality systems.
\newblock In {\em 2019 IEEE Conference on Virtual Reality and 3D User
  Interfaces (VR)}, pp. 1060--1061, 2019. doi: {{%
10\hspace{.1pt}\discretionary{.}{%
}{.}\hspace{.4pt}1109\discretionary{/}{%
}{/}VR\hspace{.1pt}\discretionary{.}{%
}{.}\hspace{.4pt}2019\hspace{.1pt}\discretionary{.}{%
}{.}\hspace{.4pt}8797901}}


\bibitem{Lu.2021.Itext}
X.~Lu, D.~Yu, H.-N. Liang, and J.~Goncalves.
\newblock itext: Hands-free text entry on an imaginary keyboard for augmented
  reality systems.
\newblock In {\em The 34th Annual ACM Symposium on User Interface Software and
  Technology}, UIST '21, p. 815–825. Association for Computing Machinery, New
  York, NY, USA, 2021. doi: {{%
10\hspace{.1pt}\discretionary{.}{%
}{.}\hspace{.4pt}1145\discretionary{/}{%
}{/}3472749\hspace{.1pt}\discretionary{.}{%
}{.}\hspace{.4pt}3474788}}


\bibitem{hands_free_text_input}
X.~Lu, D.~Yu, H.-N. Liang, W.~Xu, Y.~Chen, X.~Li, and K.~Hasan.
\newblock Exploration of hands-free text entry techniques for virtual reality.
\newblock In {\em 2020 IEEE International Symposium on Mixed and Augmented
  Reality (ISMAR)}, p.~6, 2020. doi: {{%
10\hspace{.1pt}\discretionary{.}{%
}{.}\hspace{.4pt}1109\discretionary{/}{%
}{/}ISMAR50242\hspace{.1pt}\discretionary{.}{%
}{.}\hspace{.4pt}2020\hspace{.1pt}\discretionary{.}{%
}{.}\hspace{.4pt}00061}}


\bibitem{4343891}
P.~Lukowicz, A.~Timm-Giel, M.~Lawo, and O.~Herzog.
\newblock Wearit@work: Toward real-world industrial wearable computing.
\newblock {\em IEEE Pervasive Computing}, 6(4):8--13, 2007. doi: {{%
10\hspace{.1pt}\discretionary{.}{%
}{.}\hspace{.4pt}1109\discretionary{/}{%
}{/}MPRV\hspace{.1pt}\discretionary{.}{%
}{.}\hspace{.4pt}2007\hspace{.1pt}\discretionary{.}{%
}{.}\hspace{.4pt}89}}


\bibitem{Mine95virtualenvironment}
M.~Mine.
\newblock Virtual environment interaction techniques.
\newblock Technical report, UNC Chapel Hill CS Dept, 1995.

\bibitem{729527}
J.~Ockerman and A.~Pritchett.
\newblock Preliminary investigation of wearable computers for task guidance in
  aircraft inspection.
\newblock In {\em Digest of Papers. Second International Symposium on Wearable
  Computers (Cat. No.98EX215)}, pp. 33--40, 1998. doi: {{%
10\hspace{.1pt}\discretionary{.}{%
}{.}\hspace{.4pt}1109\discretionary{/}{%
}{/}ISWC\hspace{.1pt}\discretionary{.}{%
}{.}\hspace{.4pt}1998\hspace{.1pt}\discretionary{.}{%
}{.}\hspace{.4pt}729527}}


\bibitem{Pandey.2021.Speech}
L.~Pandey, K.~Hasan, and A.~S. Arif.
\newblock Acceptability of speech and silent speech input methods in private
  and public.
\newblock In {\em Proceedings of the 2021 CHI Conference on Human Factors in
  Computing Systems}, CHI '21, p.~13. Association for Computing Machinery, New
  York, NY, USA, 2021. doi: {{%
10\hspace{.1pt}\discretionary{.}{%
}{.}\hspace{.4pt}1145\discretionary{/}{%
}{/}3411764\hspace{.1pt}\discretionary{.}{%
}{.}\hspace{.4pt}3445430}}


\bibitem{10.1145/1054972.1055013}
B.~A. Po, B.~D. Fisher, and K.~S. Booth.
\newblock Comparing cursor orientations for mouse, pointer, and pen
  interaction.
\newblock In {\em Proceedings of the SIGCHI Conference on Human Factors in
  Computing Systems}, CHI '05, p.~10. Association for Computing Machinery, New
  York, NY, USA, 2005. doi: {{%
10\hspace{.1pt}\discretionary{.}{%
}{.}\hspace{.4pt}1145\discretionary{/}{%
}{/}1054972\hspace{.1pt}\discretionary{.}{%
}{.}\hspace{.4pt}1055013}}


\bibitem{Marco_dwell}
M.~Porta and M.~Turina.
\newblock <i>eye</i>-s: A full-screen input modality for pure eye-based
  communication.
\newblock In {\em Proceedings of the 2008 symposium on Eye tracking research
  $\&$ applications}, ETRA '08, p.~8. Association for Computing Machinery, New
  York, NY, USA, 2008. doi: {{%
10\hspace{.1pt}\discretionary{.}{%
}{.}\hspace{.4pt}1145\discretionary{/}{%
}{/}1344471\hspace{.1pt}\discretionary{.}{%
}{.}\hspace{.4pt}1344477}}


\bibitem{quinn_asian_2007}
E.~Quinn, I.~S.~P. Nation, and S.~Millett.
\newblock Asian and {Pacific} speed readings for {ESL} learners.
\newblock {\em ELI Occasional Publication}, 24:74, 2007.

\bibitem{10.1145/642611.642632}
R.~W. Soukoreff and I.~S. MacKenzie.
\newblock Metrics for text entry research: An evaluation of msd and kspc, and a
  new unified error metric.
\newblock In {\em Proceedings of the SIGCHI Conference on Human Factors in
  Computing Systems}, CHI '03, p.~8. Association for Computing Machinery, New
  York, NY, USA, 2003. doi: {{%
10\hspace{.1pt}\discretionary{.}{%
}{.}\hspace{.4pt}1145\discretionary{/}{%
}{/}642611\hspace{.1pt}\discretionary{.}{%
}{.}\hspace{.4pt}642632}}


\bibitem{Marco_selection_text}
M.~Speicher, A.~M. Feit, P.~Ziegler, and A.~Kr\"{u}ger.
\newblock Selection-based text entry in virtual reality.
\newblock In {\em Proceedings of the 2018 CHI Conference on Human Factors in
  Computing Systems}, CHI '18, p.~13. Association for Computing Machinery, New
  York, NY, USA, 2018. doi: {{%
10\hspace{.1pt}\discretionary{.}{%
}{.}\hspace{.4pt}1145\discretionary{/}{%
}{/}3173574\hspace{.1pt}\discretionary{.}{%
}{.}\hspace{.4pt}3174221}}


\bibitem{Vergari.2021.SocialEnv}
M.~Vergari, T.~Kojić, F.~Vona, F.~Garzotto, S.~Möller, and J.-N.
  Voigt-Antons.
\newblock Influence of interactivity and social environments on user experience
  and social acceptability in virtual reality.
\newblock In {\em 2021 IEEE Virtual Reality and 3D User Interfaces (VR)}, pp.
  695--704, 2021. doi: {{%
10\hspace{.1pt}\discretionary{.}{%
}{.}\hspace{.4pt}1109\discretionary{/}{%
}{/}VR50410\hspace{.1pt}\discretionary{.}{%
}{.}\hspace{.4pt}2021\hspace{.1pt}\discretionary{.}{%
}{.}\hspace{.4pt}00096}}


\bibitem{wei_reading_2020}
C.~Wei, D.~Yu, and T.~Dingier.
\newblock Reading on {3D} {Surfaces} in {Virtual} {Environments}.
\newblock In {\em 2020 {IEEE} {Conference} on {Virtual} {Reality} and {3D}
  {User} {Interfaces} ({VR})}, p.~8. IEEE, Atlanta, GA, USA, Mar. 2020.
\newblock ISSN: 2642-5254. doi: {{%
10\hspace{.1pt}\discretionary{.}{%
}{.}\hspace{.4pt}1109\discretionary{/}{%
}{/}VR46266\hspace{.1pt}\discretionary{.}{%
}{.}\hspace{.4pt}2020\hspace{.1pt}\discretionary{.}{%
}{.}\hspace{.4pt}00095}}


\bibitem{Jacob_ART}
J.~O. Wobbrock, L.~Findlater, D.~Gergle, and J.~J. Higgins.
\newblock The aligned rank transform for nonparametric factorial analyses using
  only anova procedures.
\newblock In {\em Proceedings of the SIGCHI Conference on Human Factors in
  Computing Systems}, CHI '11, p.~4. Association for Computing Machinery, New
  York, NY, USA, 2011. doi: {{%
10\hspace{.1pt}\discretionary{.}{%
}{.}\hspace{.4pt}1145\discretionary{/}{%
}{/}1978942\hspace{.1pt}\discretionary{.}{%
}{.}\hspace{.4pt}1978963}}


\bibitem{xu_pointing_2019}
W.~Xu, H.-N. Liang, A.~He, and Z.~Wang.
\newblock Pointing and {Selection} {Methods} for {Text} {Entry} in {Augmented}
  {Reality} {Head} {Mounted} {Displays}.
\newblock In {\em 18th {IEEE} {International} {Symposium} on {Mixed} and
  {Augmented} {Reality} ({ISMAR})}, pp. 419--428. IEEE Computer Society,
  Beijing, China, 2019. doi: {{%
10\hspace{.1pt}\discretionary{.}{%
}{.}\hspace{.4pt}1109\discretionary{/}{%
}{/}ISMAR\hspace{.1pt}\discretionary{.}{%
}{.}\hspace{.4pt}2019\hspace{.1pt}\discretionary{.}{%
}{.}\hspace{.4pt}00026}}


\bibitem{xu.2019.dmove}
W.~Xu, H.-N. Liang, Y.~Zhao, D.~Yu, and D.~Monteiro.
\newblock Dmove: Directional motion-based interaction for augmented reality
  head-mounted displays.
\newblock In {\em Proceedings of the 2019 CHI Conference on Human Factors in
  Computing Systems}, CHI '19, p.~14. Association for Computing Machinery, New
  York, NY, USA, 2019. doi: {{%
10\hspace{.1pt}\discretionary{.}{%
}{.}\hspace{.4pt}1145\discretionary{/}{%
}{/}3290605\hspace{.1pt}\discretionary{.}{%
}{.}\hspace{.4pt}3300674}}


\bibitem{Ringtext}
W.~Xu, H.-N. Liang, Y.~Zhao, T.~Zhang, D.~Yu, and D.~Monteiro.
\newblock Ringtext: Dwell-free and hands-free text entry for mobile
  head-mounted displays using head motions.
\newblock {\em IEEE Transactions on Visualization and Computer Graphics},
  25(5):1991--2001, 2019. doi: {{%
10\hspace{.1pt}\discretionary{.}{%
}{.}\hspace{.4pt}1109\discretionary{/}{%
}{/}TVCG\hspace{.1pt}\discretionary{.}{%
}{.}\hspace{.4pt}2019\hspace{.1pt}\discretionary{.}{%
}{.}\hspace{.4pt}2898736}}


\bibitem{xu_withhands_2022}
W.~Xu, X.~Meng, K.~Yu, S.~Sarcar, and H.-N. Liang.
\newblock Evaluation of text selection techniques in virtual reality
  head-mounted displays.
\newblock In {\em 21st {IEEE} {International} {Symposium} on {Mixed} and
  {Augmented} {Reality} ({ISMAR})}, p.~10. IEEE Computer Society, Singapore,
  2022.

\bibitem{Voice_input}
N.~Yankelovich, G.-A. Levow, and M.~Marx.
\newblock Designing speechacts: Issues in speech user interfaces.
\newblock In {\em Proceedings of the SIGCHI Conference on Human Factors in
  Computing Systems}, CHI '95, p.~8. ACM Press/Addison-Wesley Publishing Co.,
  USA, 1995. doi: {{%
10\hspace{.1pt}\discretionary{.}{%
}{.}\hspace{.4pt}1145\discretionary{/}{%
}{/}223904\hspace{.1pt}\discretionary{.}{%
}{.}\hspace{.4pt}223952}}


\bibitem{tap_dwel_gesture}
C.~Yu, Y.~Gu, Z.~Yang, X.~Yi, H.~Luo, and Y.~Shi.
\newblock Tap, dwell or gesture? exploring head-based text entry techniques for
  hmds.
\newblock In {\em Proceedings of the 2017 CHI Conference on Human Factors in
  Computing Systems}, CHI '17, p.~10. Association for Computing Machinery, New
  York, NY, USA, 2017. doi: {{%
10\hspace{.1pt}\discretionary{.}{%
}{.}\hspace{.4pt}1145\discretionary{/}{%
}{/}3025453\hspace{.1pt}\discretionary{.}{%
}{.}\hspace{.4pt}3025964}}


\bibitem{Pizzatext}
D.~Yu, K.~Fan, H.~Zhang, D.~Monteiro, W.~Xu, and H.-N. Liang.
\newblock Pizzatext: Text entry for virtual reality systems using dual
  thumbsticks.
\newblock {\em IEEE Transactions on Visualization and Computer Graphics},
  24(11):2927--2935, 2018. doi: {{%
10\hspace{.1pt}\discretionary{.}{%
}{.}\hspace{.4pt}1109\discretionary{/}{%
}{/}TVCG\hspace{.1pt}\discretionary{.}{%
}{.}\hspace{.4pt}2018\hspace{.1pt}\discretionary{.}{%
}{.}\hspace{.4pt}2868581}}


\bibitem{yu.2018.targetselection}
D.~Yu, H.-N. Liang, F.~Lu, V.~Nanjappan, K.~Papangelis, and W.~Wang.
\newblock Target selection in head-mounted display virtual reality
  environments.
\newblock {\em Journal of Universal Computer Science}, 24(9):1271--1243, 2018.

\bibitem{pointingstudy1}
D.~Yu, H.-N. Liang, X.~Lu, K.~Fan, and B.~Ens.
\newblock Modeling endpoint distribution of pointing selection tasks in virtual
  reality environments.
\newblock {\em ACM Trans. Graph.}, 38(6):13, nov 2019. doi: {{%
10\hspace{.1pt}\discretionary{.}{%
}{.}\hspace{.4pt}1145\discretionary{/}{%
}{/}3355089\hspace{.1pt}\discretionary{.}{%
}{.}\hspace{.4pt}3356544}}


\bibitem{DepthMove}
D.~Yu, H.-N. Liang, X.~Lu, T.~Zhang, and W.~Xu.
\newblock Depthmove: Leveraging head motions in the depth dimension to interact
  with virtual reality head-worn displays.
\newblock In {\em 2019 IEEE International Symposium on Mixed and Augmented
  Reality (ISMAR)}, pp. 103--114, 2019. doi: {{%
10\hspace{.1pt}\discretionary{.}{%
}{.}\hspace{.4pt}1109\discretionary{/}{%
}{/}ISMAR\hspace{.1pt}\discretionary{.}{%
}{.}\hspace{.4pt}2019\hspace{.1pt}\discretionary{.}{%
}{.}\hspace{.4pt}00\discretionary{%
}{-}{-}20}}


\bibitem{yu.2021.gaze}
D.~Yu, X.~Lu, R.~Shi, H.-N. Liang, T.~Dingler, E.~Velloso, and J.~Goncalves.
\newblock Gaze-supported 3d object manipulation in virtual reality.
\newblock In {\em Proceedings of the 2021 CHI Conference on Human Factors in
  Computing Systems}, CHI '21, p.~13. Association for Computing Machinery, New
  York, NY, USA, 2021. doi: {{%
10\hspace{.1pt}\discretionary{.}{%
}{.}\hspace{.4pt}1145\discretionary{/}{%
}{/}3411764\hspace{.1pt}\discretionary{.}{%
}{.}\hspace{.4pt}3445343}}


\bibitem{10.1145/2702123.2702305}
X.~S. Zheng, C.~Foucault, P.~Matos~da Silva, S.~Dasari, T.~Yang, and S.~Goose.
\newblock Eye-wearable technology for machine maintenance: Effects of display
  position and hands-free operation.
\newblock In {\em Proceedings of the 33rd Annual ACM Conference on Human
  Factors in Computing Systems}, CHI '15, p.~10. Association for Computing
  Machinery, New York, NY, USA, 2015. doi: {{%
10\hspace{.1pt}\discretionary{.}{%
}{.}\hspace{.4pt}1145\discretionary{/}{%
}{/}2702123\hspace{.1pt}\discretionary{.}{%
}{.}\hspace{.4pt}2702305}}


\end{thebibliography}

\end{document}